\documentclass[aps, prl, reprint, superscriptaddress, preprintnumbers]{revtex4-1}

\usepackage{color}
\usepackage{graphicx}
\usepackage{dcolumn}
\usepackage{bm}
\usepackage{color}

\begin{document}


\title{Extension of the Coherence Time by Generating MW Dressed States in a Single NV Centre in Diamond}



\author{H. Morishita}
\email[]{h-mori@scl.kyoto-u.ac.jp}
\affiliation{These authors equally contributed to this work.}
\affiliation{Institute for Chemical Research, Kyoto University, 611-0011, Japan}

\author{T. Tashima}
\email[]{tashima.toshiyuki.5e@kyoto-u.ac.jp} 
\affiliation{These authors equally contributed to this work.}
\affiliation{Department of Electronic Science and Engineering, Kyoto University,
	615-8510 Kyoto, Japan}

\author{D. Mima}
\affiliation{Institute for Chemical Research, Kyoto University, 611-0011, Japan}

\author{H. Kato}
\affiliation{Energy Technology Research Institute, National Institute of Advanced Industrial Science and Technology (AIST), Ibaraki 305-8568, Japan}

\author{T. Makino}
\affiliation{Energy Technology Research Institute, National Institute of Advanced Industrial Science and Technology (AIST), Ibaraki 305-8568, Japan}

\author{S. Yamasaki}
\affiliation{Energy Technology Research Institute, National Institute of Advanced Industrial Science and Technology (AIST), Ibaraki 305-8568, Japan}

\author{M. Fujiwara}
\affiliation{Institute for Chemical Research, Kyoto University, 611-0011, Japan}

\author{N. Mizuochi}
\email[]{mizuochi@scl.kyoto-u.ac.jp}
\affiliation{Institute for Chemical Research, Kyoto University, 611-0011, Japan}


\date{\today}

\begin{abstract}
Nitrogen-vacancy (NV) centres in diamond hold promise in quantum sensing applications. A major interest in them is an enhancement of their sensitivity by the extension of the coherence time ($T_2$). In this report, we experimentally generated more than four dressed states in a single NV centre in diamond based on Autler-Townes splitting (ATS). We also observed the extension of the coherence time to $T_2 \sim$ 1.5 ms which is more than two orders of magnitude longer than that of the undressed states. As an example of a quantum application using these results we propose a protocol of quantum sensing, which shows more than an order of magnitude enhancement in the sensitivity.
\end{abstract}


\maketitle

A single spin of a nitrogen-vacancy (NV) centre in diamond has a long coherence time ($T_2$) under ambient conditions, and hence it is a promising candidate for classical- and quantum-sensing applications~\cite{TaylorNP08,MaurerNP10,McGuinnessNNanoT11,KucskoNat13,SageNat13,FazhanNP14,AjoyPRX15,ZhuNature11,KuboPRL11, MatsuzakiPRL15, UndenPRL16,ZaiserNC16,MatsuzakiPRA16}. Various approaches to enhance the sensitivity of the NV based sensors have been experimentally demonstrated so far. For example, hybrid quantum sensors, which are inspired by quantum memory effects~\cite{ZaiserNC16,MatsuzakiPRA16}, are of great concern. They use nuclear spins of carbon or nitrogen around the NV centre as quantum memories. If paramagnetic impurities such as substitutional nitrogen (P1) centres and $^{13}$C nuclear spins are sufficiently suppressed~\cite{WangPRB13}, we can keep $T_2$ long and the sensitivity can be enhanced~\cite{BalasubramanianNM09}. While, when we consider the situation to increase the number of nearby nuclear spins as quantum memories to enhance the sensitivity more, $T_2$ of both the electron and nuclear spins become shorter~\cite{MizuochiPRB09}. Here, we focus on the microwave (MW) dressed state based on Autler-Townes splitting (ATS)~\cite{LauchtNatNanotech17,ATS,ATSBook, HeJAP92, MansonJL10,YanNC13}. Using ATS, a large number of the dressed states can be generated~\cite{LauchtNatNanotech17}. It has also been reported that $T_2$ of the MW dressed states is longer than that of undressed states~\cite{LauchtNatNanotech17}. 

In this report, we experimentally demonstrate the generation of the MW dressed states in the single NV centre in diamond by ATS at ambient conditions in order to analyse fundamental phenomena. Next, we show the extension of $T_2$ under the generation of the dressed states. Finally, we propose a protocol for the quantum sensing with the dressed states for AC-field sensing and numerically estimate the sensitivity as functions of the number of the dressed states and $T_2$.

\begin{figure}
	\includegraphics[width=8cm,clip]{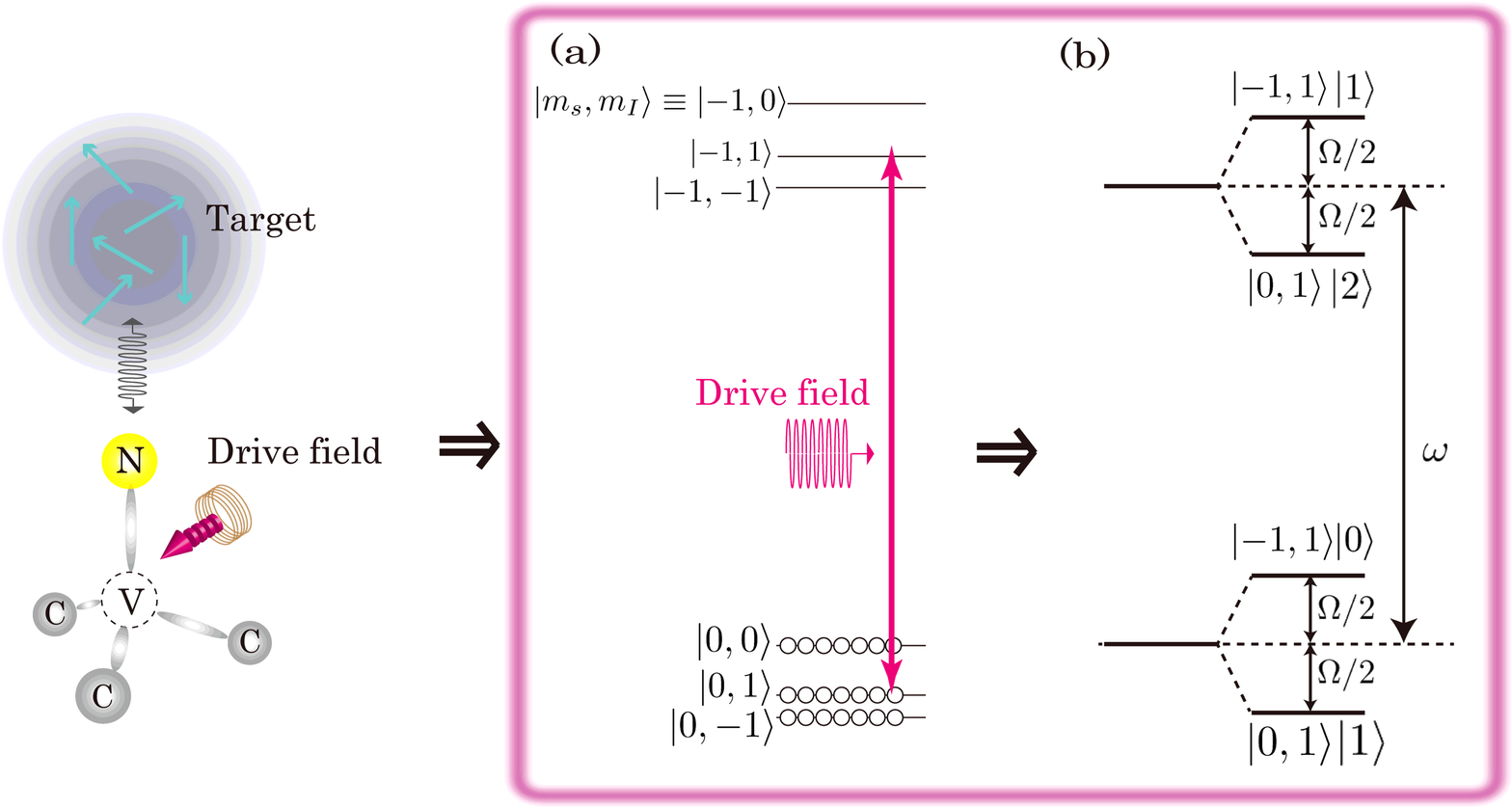}
	\caption{
	(a) Energy diagram of the NV centre under irradiation of a weak drive field. (b) Dressed energy level coupling with a mode of the drive field.
	}
	\label{figSchematics}
\end{figure}

\section{MECHANISM OF AUTLER-TOWNES SPLITTING}
Dressed states based on the ATS are experimentally generated using the single NV centre in diamond by irradiation of an MW drive field. Figure~\ref{figSchematics}(a) shows the energy level of the NV electron spin coupled with the $^{14}$N nuclear spin of the NV centre, where $\left|m_s, m_I\right>$ is defined as the electron and the $^{14}$N nuclear spin of the NV centre, respectively. After laser illumination, the NV centre is equally polarised in $\left|0,0\right>$, $\left|0,1\right>$, and $\left|0,-1\right>$ depicted by the open circles under the application of a static magnetic field ($B_0$). Figure~\ref{figSchematics}(a) also depicts the irradiation of an unperturbed drive field whose frequency is close to a resonant frequency of a transition between $\left|0,1\right>$ and $\left|-1,1\right>$. When the drive field is considered as a classical mw mode, the NV centre can be coupled to the mode of the drive field. Then, each $\left|0,1\right>$ and $\left|-1,1\right>$ is split into two levels described in Fig.~\ref{figSchematics}(b). It should be noted that Fig.~\ref{figSchematics}(b) depicts the minimum number of dressed states by the ATS as an example. Thus, Fig.~\ref{figSchematics}(b) depicts the generation of four dressed states of $\left| 0,1 \right> \left|1 \right>$, $\left|-1,1\right>\left|0\right>$, $\left|0,1\right>\left|2\right>$, and $\left|-1,1\right>\left|1\right>$ in the presence of coupling between the NV centre ($\left|0,1\right>$ and $\left|-1,1\right>$) and the mode of the drive field ($\left|0\right>$, $\left|1\right>$, and $\left|2\right>$). This phenomenon is called (weak) ATS. Figure~\ref{figSchematics}(b) also shows the energy levels of the dressed states which are characterised by the Rabi frequency of an NV electron spin ($\Omega$) and frequency of the drive field ($\omega$), and its spectrum is given by the following equation~\cite{Mollow1969}:
\begin{eqnarray}
	f(\nu) &=& \frac{\frac{1}{4}\kappa}{\left(\nu-\omega\right)^2 +\frac{1}{4}\kappa^2} + \frac{\frac{3}{16}\kappa}{\left(\nu-\omega-\Omega\right)^2 + \frac{9}{16}\kappa^2} \nonumber \\
	&& +  \frac{\frac{3}{16}\kappa}{\left(\nu-\omega+\Omega\right)^2 + \frac{9}{16}\kappa^2},
	\label{EqATS}
\end{eqnarray}
where $\nu$ is the incident probe-frequency. $\kappa$ is the inverse of the dephasing time. $\Delta \omega = \omega_0 - \omega$ . Here $\omega_0$ means the resonant frequency of an NV electron spin. The first term of Eq.~(\ref{EqATS}) shows the resonant frequency depends on just $\omega$ and the second and third terms of Eq.~(\ref{EqATS})  show the resonant frequencies depend on not only $\omega$ but also $\Omega$. Thus, Eq.~(\ref{EqATS})  is described as a Mollow-triplet spectrum. In the case of $\Delta \omega = 0$, the resonant frequency of the central peak, which is described by the first term in Eq.~(\ref{EqATS}), does not depend on the $\Omega$, while the side peaks, which are described by the second and third terms in Eq.~(\ref{EqATS}), have linear dependences on $\Omega$. In the case of $\Delta \omega \neq 0$, the resonant frequency of the central peak depends on ∆ω according to the following relation~\cite{Wei1994}: $\Omega = \sqrt{ \Omega_0^2 + (\Delta \omega)^2}$, where $\Omega_0$ is the Rabi frequency in the on-resonance condition. On the other hand, the changes in the resonant frequencies of the side peaks satisfy the following relation~\cite{Wei1994}: $\Omega = \Omega_0 \pm \sqrt{\Omega_0^2 +(\Delta \omega)^2 }$. 

\section{THE INFORMATION OF OUR SAMPLE AND EXPERIMENTAL SETUP}
\begin{figure}
 	\includegraphics[width=7cm,clip]{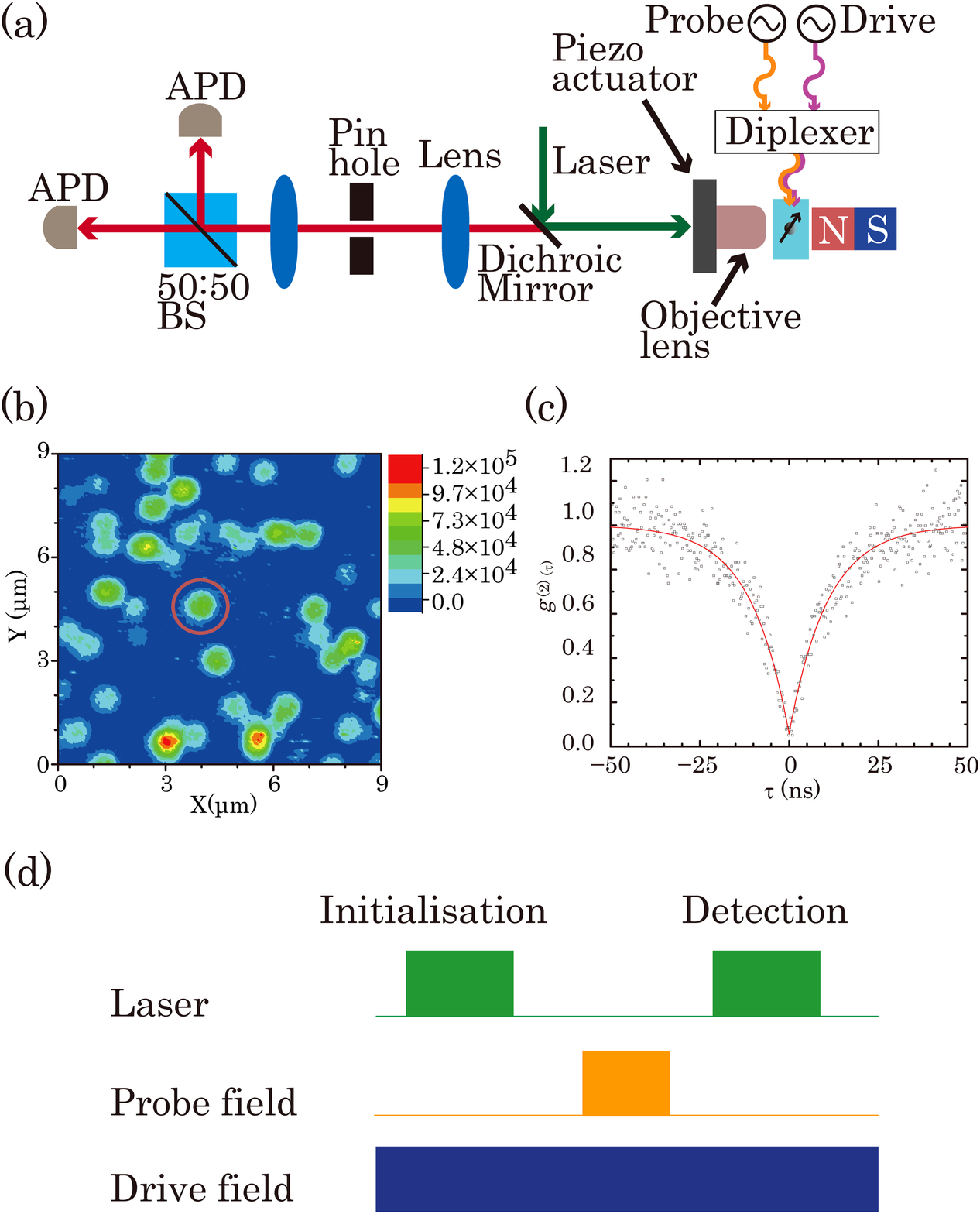}
	\caption{
		(a) Schematic of a homemade confocal microscope with an electromagnetic field (emf) irradiation system. 
		(b) Photoluminescence scanning image of the NV centres in diamond. The red circle shows the single NV centre used in this experiment. 
		(c) $g^{(2)}(\tau)$ for the NV centre. 
	}
	\label{figSetup}
\end{figure}

Figure~\ref{figSetup}(a) shows our experimental setup (See Methods). The sample is a high-temperature high-pressure (HTHP) type IIa (111) diamond. The second-order autocorrelation function, $g^{(2)}(\tau)$, was measured using the Hanbury-Brown-Twiss (HBT) setup~\cite{BerPRL15}  to confirm whether NV centre indicated by the circle depicted in Fig.~\ref{figSetup}(b) is a single centre or not. The power of the 532-nm laser is 100 $\mu$W. Figure~\ref{figSetup}(c) shows the measured $g^{(2)}(0)$ is $\sim$ 0.1. Therefore, this NV centre is a single centre.

In our experiment, we chose an NV centre that is weakly coupled to other nuclear spins (e.g., $^{13}$C nuclear spin). We measured the optically detected magnetic resonance (ODMR) spectrum with a 1-$\mu$s pulsed laser by sweeping the frequency of a 5.5-$\mu$s pulsed probe MW (P$_\mathrm{mw}$) pulse ($\pi$ pulse) depicted top of Fig.~\ref{figPowerDepend}. In Fig.~\ref{figPowerDepend}(a), the ODMR spectrum has three dips with 2.1 MHz splitting, which corresponds to the hyperfine splitting of the $^{14}$N nuclear spin of the NV centre~\cite{SteinerPRB10}.

\section{EXPERIMENTAL GENERATION OF DRESSED STATE BY ATS}
\begin{figure}
	\includegraphics[width = 7cm, clip]{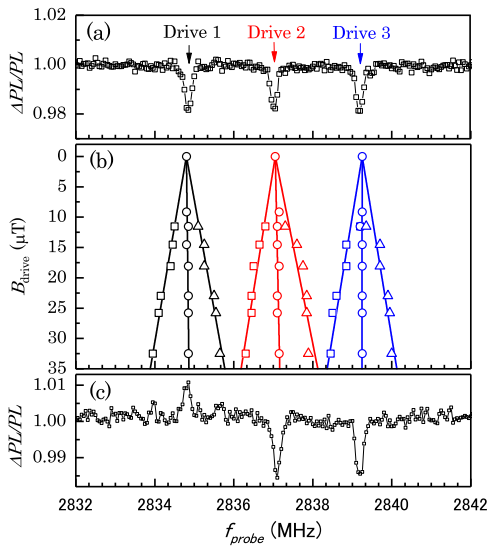}
	\caption{
		(a) ODMR spectrum without any drive fields. 
		(b) Resonant frequencies as a function of the strength of the drive field ($B_\mathrm{drive}$). Black, red, and blue plots show the changes in resonant frequencies under the irradiation of Dmw frequencies of 2834.75 MHz (D$_\mathrm{mw}$1), 2837.05 MHz (D$_\mathrm{mw}$2), and 2839.18 MHz (D$_\mathrm{mw}$3), respectively. Solid lines are fitted for each resonant frequency. 
		(c) ODMR spectrum under a Dmw at a frequency of 2834.75 MHz and a D$_\mathrm{mw}$ power of 33 $\mu$T. We can observe the Mollow triplet, which we call ATS.
	} 
	\label{figPowerDepend}
\end{figure}

First, we measured the change in the dressed-state resonant frequencies by changing the power of continuous drive MW (D$_\mathrm{mw}$) with pulse sequence depicted in the top of Fig.\ref{figPowerDepend}. These experiments use the three D$_\mathrm{mw}$ frequencies of 2834.75 MHz (D$_\mathrm{mw}$1), 2837.05 MHz (D$_\mathrm{mw}$2), and 2839.18 MHz (D$_\mathrm{mw}$3) to generate dressed states. The results are shown in Fig.~\ref{figPowerDepend}(b). The signals for each D$_\mathrm{mw}$ frequency split into three above $\sim$ 10 $\mu$T. 

Here we focus on the 2834.75 MHz of D$_\mathrm{mw}$1. It should be noted that all three D$_\mathrm{mw}$ frequencies have the same dependences on the power of the continuous D$_\mathrm{mw}$. The ODMR spectrum under continuous irradiation at D$_\mathrm{mw}$1 with the power of 33 $\mu$T is shown in Fig.~\ref{figPowerDepend}(c). It shows the increase of PL intensity was observed around 2835 MHz with continuous irradiation of D$_\mathrm{mw}$1 (Fig.~\ref{figPowerDepend}(c)) while the decrease of PL intensity was observed around 2835 MHz without continuous irradiation of D$_\mathrm{mw}$1 (see Fig.~\ref{figPowerDepend}(a)). The peaks are inverted, and the reason may come from the pulse sequence depicted in the top of Fig.~\ref{figPowerDepend}. It shows the pulsed laser and the continuous D$_\mathrm{mw}$ field are simultaneously applied to the NV centre in the initialisation process. When the NV centre can be initialised into $\left|-1\right>$ by the pulsed laser and continuous D$_\mathrm{mw}$ field, the increase of PL intensity may be observed at magnetic resonance conditions of the NV electron spin as discussed in Ref.~\cite{KehayiasPRB14}. In addition, Fig.~\ref{figPowerDepend}(c) shows the ODMR spectrum at $\sim$ 2834.75 MHz splits into three peaks under the irradiation of the D$_\mathrm{mw}$. Figure~\ref{figPowerDepend}(b) shows that the resonant frequencies of the dressed states as a function of $B_\mathrm{drive}$. The solid lines show the linear fitting for each observed data. The absolute values of these slopes in Fig.~\ref{figPowerDepend}(b) agree well with the gyromagnetic ratio of the NV electron spin ($\gamma_\mathrm{NV}$)\cite{Doherty13}, so that means the resonant frequencies of side peaks are linearly proportional to the Rabi frequencies of the NV electron spin. This result is in agreement with the theory of the change of resonant frequencies according to the second and third terms in Eq.Eq.~(\ref{EqATS}) with $\Omega = \gamma_\mathrm{NV} B_\mathrm{drive}$. 

\begin{figure}
	\includegraphics[width=7cm,clip]{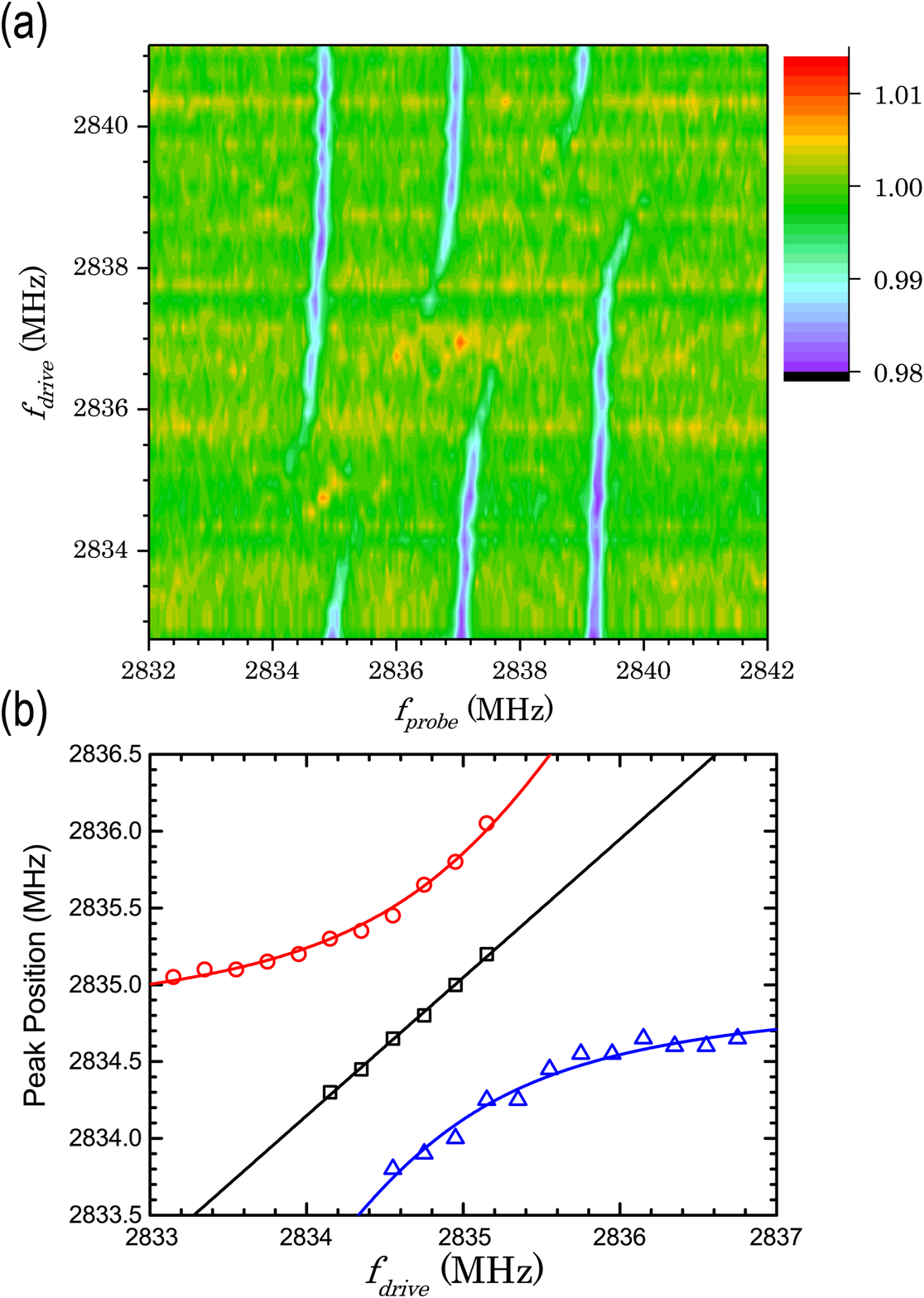}
	\caption{
		(a) $\Delta$PL/PL intensity plots as functions of the D$_\mathrm{mw}$ vs P$_\mathrm{mw}$ frequencies. 
		(b) Resonant frequencies of dressed states as a function of the Dmw frequencies. The black line shows a linear fitting with $\Omega = \sqrt{ \Omega_0^2 + (\Delta \omega)^2}$. Red and blue solid lines show fittings with $\Omega = \Omega_0 \pm \sqrt{\Omega_0^2 +(\Delta \omega)^2 }$.  
	}
	\label{figFreqDepend}
\end{figure}

Next, we measured the change in the resonant frequencies of the dressed states by changing the continuous D$_\mathrm{mw}$ frequency while fixing the D$_\mathrm{mw}$ power at 33 $\mu$T with the pulse sequence depicted in the top of Fig.~\ref{figFreqDepend}. The D$_\mathrm{mw}$ frequency was changed by the step of 0.2 MHz. The result is shown in Fig.~\ref{figFreqDepend}(a). It shows that he dressed states are generated when the D$_\mathrm{mw}$ frequencies are close to resonant frequencies of the NV electron spin.

Figure~\ref{figFreqDepend}(b) illustrates the dressed-state resonant frequencies as a function of the D$_\mathrm{mw}$ frequency around the centre of 2835 MHz to understand more details of the results depicted in Fig.~\ref{figFreqDepend}(a). The black squares in Fig.~\ref{figFreqDepend}(b) show the change of resonant frequencies of the centre peaks as a function of D$_\mathrm{mw}$ frequency. The dependence can be fitted by a linear function shown in solid black line in Fig.~\ref{figFreqDepend}(b), and hence this is in good agreement with the theoretical prediction of $\Omega = \sqrt{ \Omega_0^2 + (\Delta \omega)^2}$ described in Eq.~(\ref{EqATS}. The red circles and the blue triangles in Fig.~\ref{figFreqDepend}(b) also show the resonant frequencies of low- and high-resonant frequencies of the side peaks, respectively. Their resonant frequencies can be fitted by using the relation of the $\Omega = \Omega_0 \pm \sqrt{\Omega_0^2 +(\Delta \omega)^2 }$~\cite{Wei1994}. Thus, all results are consistent with the theoretical prediction in ATS, demonstrating the generation of more than four dressed states by the ATS. 

\section{COHERENCE TIME OF DRESSED STATES}
\begin{figure}
	\includegraphics[width=6cm,clip]{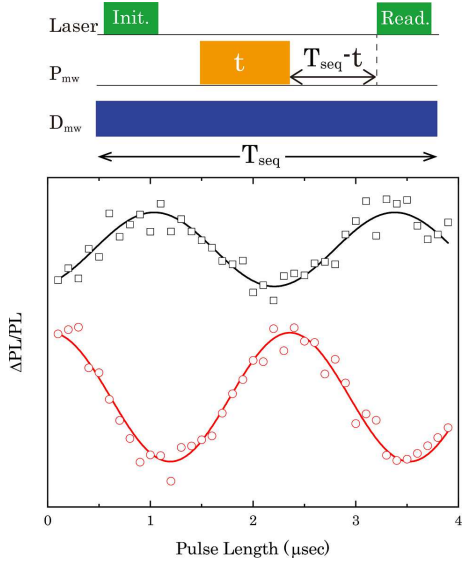}
	\caption{
		(Top) Pulse sequence to observe Rabi oscillations of the dressed state and undressed NV electron spin. (Bottom) Black and red plots show the results of the Rabi oscillations of the dressed state and the undressed NV electron spin, respectively. They are fitted by sinusoidal curve described by black and red solid lines.
	}
	\label{figRabi}
\end{figure}

First, we show whether the magnetic moment of the dressed states is the same with that of the undressed NV electron spin or not, since the dressed states were generated due to the coupling between the NV electron spins and mode of the  D$_\mathrm{mw}$. In order to investigate the magnetic moments of the dressed and undressed states, we measured Rabi oscillations of the dressed states and the NV electron spin with the pulse sequence depicted in the top of Fig.~\ref{figRabi}. It noted that while the Rabi oscillation of the dressed state was measured with continuous  D$_\mathrm{mw}$, the Rabi oscillation of the NV electron spin was measured without continuous  D$_\mathrm{mw}$. Moreover, we kept the pulse sequence time ($T_\mathrm{seq}$) constant adjusting interval between the pulsed  P$_\mathrm{mw}$ and the readout laser pulse depicted in Fig.~\ref{figRabi}, in order to perform the initialisation of the dressed states by the simultaneous irradiation of the pulse laser and the continuous  D$_\mathrm{mw}$ in the Rabi measurements. The bottom of Fig.~\ref{figRabi} shows the results of the Rabi measurements, and it indicates the Rabi frequency of the dressed spin states is the same with that of undressed spin states. Consequently, the magnetic moment of the NV electron spin and the dressed states are the same with each other. 

\begin{figure}
	\includegraphics[width=8cm,clip]{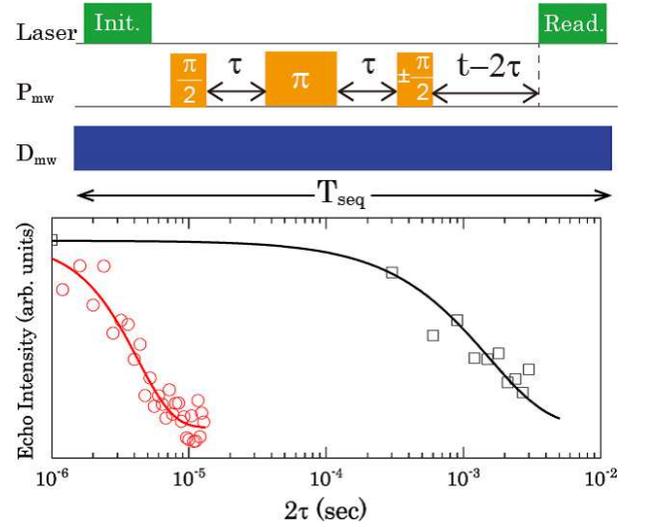}
	\caption{
		(Top) Pulse sequence to observe$T_{2\rho}$ and $T_2$ with applying a phase cycle to the final $\pi$/2 pulse. (Bottom) Black and red plots show the results of $T_{2\rho}$ and $T_2$ measurements, respectively. They are fitted by exponential decay curves described by black and red solid lines.
	}
	\label{figEcho}
\end{figure}

Next, we experimentally measured coherence time of the dressed states ($T_{2\rho}$) and coherence time of a single NV centre ($T_2$) in a $^{12}$C enriched diamond, since a coherent oscillation due to $^{13}$C nuclear spins on echo measurements~\cite{ChildressScience06} can be suppressed using the NV centre in $^{12}$C enriched diamond. The top of Fig.~\ref{figEcho} shows the pulse sequence for the $T_{2\rho}$ and $T_2$ measurements. It is noted that while $T_{2\rho}$ was measured with the continuous $D_\mathrm{mw}$ irradiation, $T_2$ was measured without continuous D$_\mathrm{mw}$ irradiation. Since the pulsed laser and continuous D$_\mathrm{mw}$ were simultaneously irradiated to the NV centre during the $T_{2\rho}$ measurements, we kept the pulse sequence time ($T_\mathrm{seq}$) constant adjusting interval between the final $\pi$/2 P$_\mathrm{mw}$ pulse and the readout laser pulse depicted in Fig.~\ref{figEcho}. Then, the dressed spin states can be initialised by simultaneous irradiation of the pulse laser and the continuous D$_\mathrm{mw}$ in the $T_{2\rho}$ measurements. Additionally, a phase cycling technique was applied to $T_{2\rho}$ measurements in order to remove common-mode noise from laser fluctuations\cite{PhamPRB16}. It is noted that the phase of the last $\pi$/2 P$_\mathrm{mw}$ pulse is indicated by $\pm$ signs depicted in the top of Fig.~\ref{figEcho}. In the case of a P$_\mathrm{mw}$ and D$_\mathrm{mw}$ strength of $\sim$ 0.43 MHz and $\sim$ 1.2 MHz, respectively, the result of $T_{2\rho}$ (black plots) and $T_2$ (red plots) measurements fitted with exponential decay curves are shown in Fig.~\ref{figEcho}. The results show that we observed a coherence time of $T_{2\rho} \sim$ 1.5 ms of the dressed states, which is more than two orders of magnitude longer than $T_2 \sim$ 4.2 μs of the undressed states. While such an extension can also be demonstrated by a dynamical decoupling technique, e.g., a Carr-Purcell-Meiboom-Gill (CPMG) sequence in the NV centres~\cite{FazhanNP14,LangeScience10,NaydenovPRB11,PhamPRB12,FarfurnikPRB15}, an extension of two orders of magnitude by the ATS is much larger than the extension of about one order of $T_2$ in the dynamical decoupling techniques~\cite{FazhanNP14,PhamPRB16,LangeScience10,NaydenovPRB11,PhamPRB12}. The extended $T_2$ by the ATS is also close to the longest $T_2$ of a single NV centre in a $^{12}$C enriched diamond~\cite{BalasubramanianNM09}.

\section{ESTIMATION OF THE SENSITIVITY OF THE QUANTUM SENSING WITH DRESSED STATES}
\begin{figure}
	\includegraphics[width=8cm,clip]{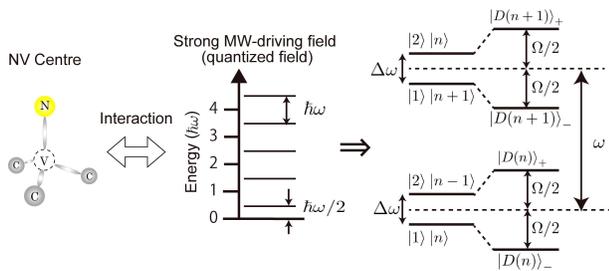}
	\caption{
		When an NV centre in diamond (left) interacts with a mode of a strong D$_\mathrm{mw}$ (centre), the NV centre coupled with the mode generates dressed states (right).
	}
	\label{figATS}
\end{figure}

We propose a quantum sensing protocol with the MW dressed states for AC magnetic field sensing and discuss the numerical estimation of the sensitivity for this sensing under a simple and ideal case. First, we shortly explain how to generate an arbitrary number of MW dressed states. Here we consider the NV centre is coupled with the mode of a strong D$_\mathrm{mw}$, as illustrated in Fig.~\ref{figATS}. The first energy is $\hbar \omega$/2, and the other states are separated by $\hbar \omega$, where $\hbar$ and $\omega$ are the reduced Planck constant and the frequency of D$_\mathrm{mw}$, respectively. If such a mode of theD$_\mathrm{mw} \left(\left|n\right>\right)$ is coupled with two states of the NV centre of $\left| 1 \right> $ and $\left| 2\right>$, dressed states of $\left| D(n)\right>_\pm$ appear. The $\left| D(n)\right>_\pm$ and their energies  $\left( E_\pm (n)\right)$ are described by the following equations~\cite{ATSBook,AtomPhoton92}:
\begin{eqnarray}
\left|D(n)\right>_\pm = c_1\left|1,n\right> \pm c_2 \left|2,n-1\right>, \\
E_\pm (n) = \left(n - \frac{1}{2}\right) \hbar \omega \pm \frac{1}{2}\hbar \Omega,
\end{eqnarray}
respectively. $c_1$ and $c_2$ are the coefficients, which satisfy $\left|c_1\right|^2 + \left|c_2\right|^2 = 1$. Since a number of photons is large in the classical mw field mode~\cite{LauchtNatNanotech17}, a large number of dressed states can be generated. Then, we discuss the sensitivity of the AC magnetic field sensing using the arbitrary number of dressed states. Our experimental demonstrations show the magnetic moments of dressed states are the same with that of undressed states and the coherence time of the dressed states is more than two orders of magnitude longer that of the undressed states. Based on the above discussion, the ratio of the sensitivities with and without the dressed states is taken by $\sqrt{MT_{2\rho}/T_2}$, where 2$M$ corresponds to the number of dressed states (Deraitls are disucssed in Supplementary Informaiton). In the case of $T_{2\rho} \sim$ 1.5 ms and $T_2 \sim$ 4.2 $\mu$s, the sensitivity is approximately enhanced 27 times with $M$ = 2. Thus, the sensitivity can be effectively enhanced by using the dressed states. In particular, our protocol is very useful for an ensemble system which has very short T2, because of two effects: ‘the extension of T2’ and ‘the addition of the sensing information measured by generated dressed states’. Although we, here, do not consider the number of operations of the integration is limited by $T_2$, our proposal opens up a new way for higher sensitivity on NV based AC magnetic field sensing. 

\section{Conclusion}
In conclusion, we have experimentally demonstrated the generation of more than four dressed states of an NV centre in diamond with irradiating the continuous D$_\mathrm{mw}$ based on ATS. Our experimental results were good agreement with the theoretical prediction. Additionally, we have proposed a new quantum sensing protocol with the dressed states for AC magnetic fields sensing. Numerical estimations show the sensitivity of the quantum sensing with the dressed states can be enhanced at least one-order of magnitude with experimentally observed $T_{2\rho}$ and $T_2$. Thus, we believe that the quantum sensing with the dressed states can be applicable for improving the sensitivity of a quantum sensing.

\textbf{Note:}
Recently, we have become aware of related works on quantum sensing with a one-time measurement based on a combination of the Mollow triplet and dynamical decoupling under sensing of a weak AC field with GHz frequencies~\cite{JoasNC17,StarkarNatCom17}. Our work has three differences: 1) the frequency range for the sensing target, 2) the effect of integration of dressed states for a higher sensitivity, and 3) robustness against environmental noises (extension of T2). Our work can realise sensing of a weak low-frequency AC field by dressed states generated by ATS.

\section{Methods}
\subsubsection{Sample Preparation}
To generate dressed states by ATS, we used high-temperature and high-pressure (HTHP) type IIa (111) diamond. After the nitrogen ($^{14}$N) was implanted into the diamond with a 30-keV accelerating energy, the sample was annealed at 750 $^{\circ}$C for 30 min for the generation of NV centres in diamond. To measure $T_2$ and $T_{2\rho}$, we used a single NV centre in a CVD-grown $^{12}$C enriched diamond layer on a type Ib (111) diamond substrate. The NV centres were generated during the growth of the diamond layer. 

\subsubsection{Home-made confocal microscope with an electromagnetic field irradiation system}
All experiments were performed by a homemade confocal microscope with an electromagnetic field (emf) irradiation system at room temperature depicted in Fig.~\ref{figSetup}(a). A 532-nm laser focused by an objective lens illuminates an NV centre in diamond. The detection system is composed of a 50:50 beam splitter (BS) and two avalanche photodiodes (APDs) in order to detect the photoluminescence and measure $g^{(2)}(\tau)$. Two high frequencies with $\sim$ 2.8 GHz irradiate to the NV centre by a by a thin copper wire with a diameter of 10 $\mu$m to manipulate the electron spin of an NV centre under the application of a static magnetic field generated by a neodymium magnet.

\section{Acknowledgments}
This work is supported by KAKENHI (No. 15H05868, 16H02088). HM is supported by a Grant-in-Aid for Young Scientists (B), Grant No. 16K17484 and by the Future Development Funding Program of Kyoto University Research Coordination Alliance.

\section{Authors Contribution}
HM, TT, and DM performed the measurements and the data analysis. HK synthesised the $^{12}$C-enriched diamond layers. All the authors contributed to the data analysis, discussion, and manuscript preparation.

\section{Additional Information}
Competing Interests: The authors declare no competing financial interests.


\newpage{}
\section{Supplementary Information}
\subsection{Estimation of the sensitivity of an NV magnetometer with mw dressed states}
\begin{figure}
	\includegraphics[width=6cm,clip]{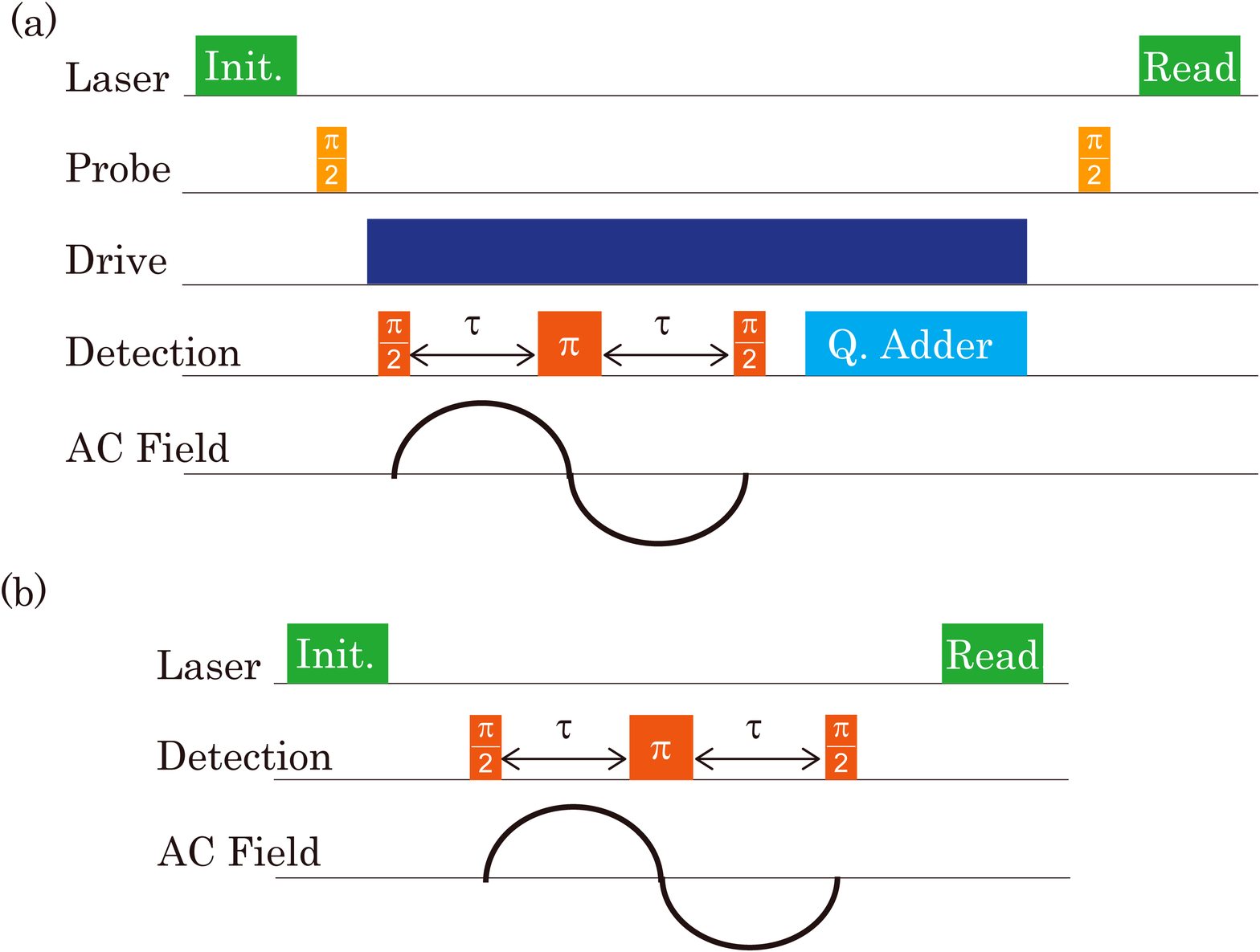}
	\caption{
		Pulse sequence to demonstrate AC magnetic field sensing (a) with and (b) without virtual quantum states.
	}
	\label{figSeq}
\end{figure}

\begin{figure}
	\includegraphics[width=5cm,clip]{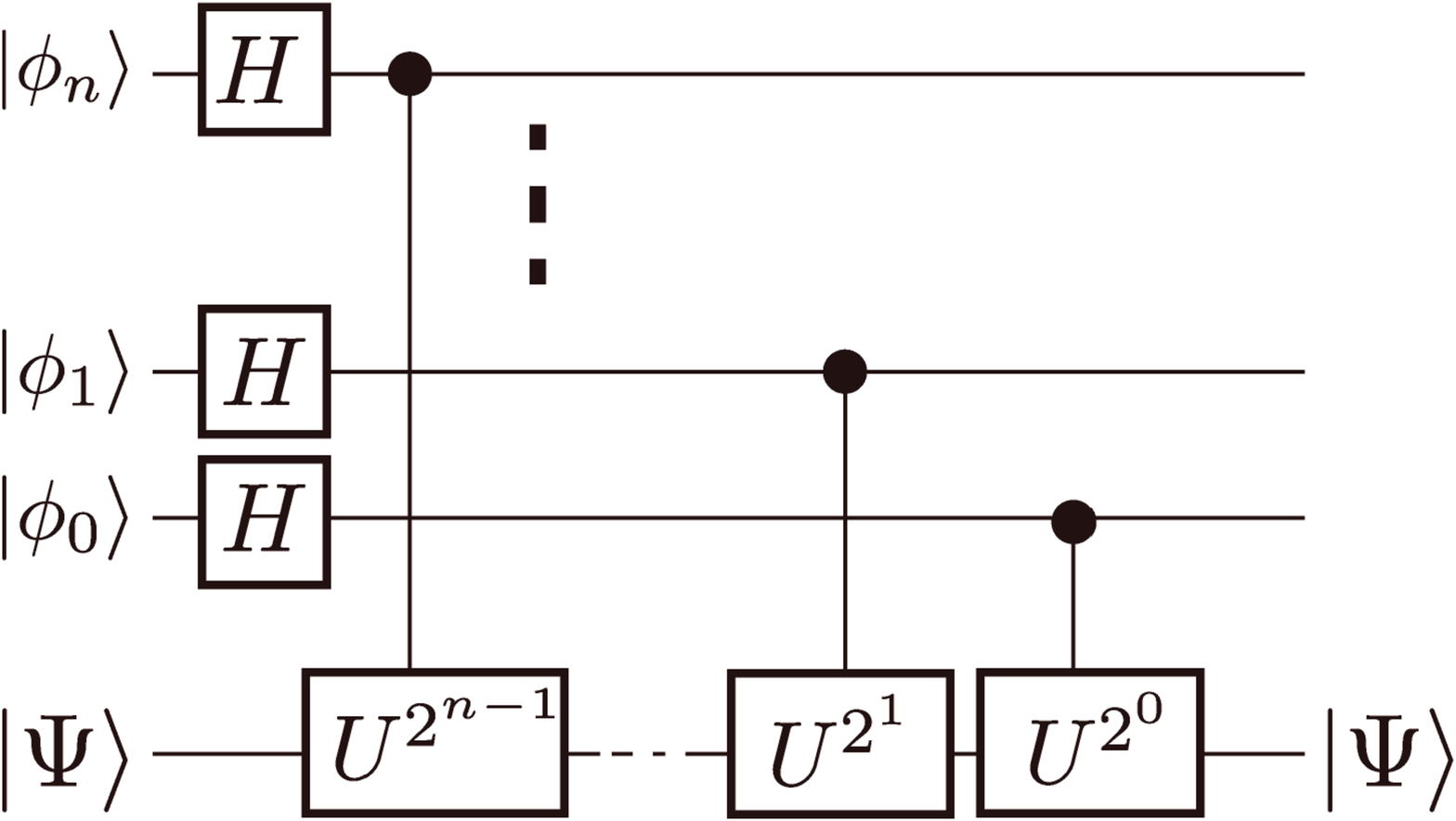}
	\caption{Quantum circuit of the quantum adder.}
	\label{figQAdder}
\end{figure}

We estimate the sensitivity of a quantum sensor with mw dressed states in the case of sensing for weak AC magnetic field. The key of our sensing is that the sensing information of the AC magnetic field in the mw dressed states can be adding. By using the fact that the dressed states have the same resonant frequencies, the number of dressed states generated for our sensing are then used in order to boost up the sensitivity of the target. Namely, we can add the information corrected at the same time using a quantum adder~\cite{QuantumAdderS}. Figure~\ref{figQAdder}(a) shows the sequence to demonstrate the quantum sensing with the dressed states for the weak AC magnetic field. This sequence is considered as a combination of generation of dressed states and conventional weak AC magnetic field sensing with a Hahn echo sequence described in Fig.~\ref{figQAdder}(b)\cite{TaylorNP08S}. The details of the sequence described in Fig.~\ref{figQAdder}(a) are followings: after the initialisation of an NV electron spin by a pulse laser, applications of a first $\pi$/ 2 pulse and pulsed strong mw-driving field generate dressed states in the NV centre. Under the generation of the dressed states, the sensing information of the AC magnetic field is stored in these states by using the Hahn echo sequence. Finally, the stored information shown in Fig.~\ref{figSeq} is corrected by a quantum adder. The quantum adder consists of a Hadamard gate $H$ and a unitary operator $U^n$ of $\left|\Psi \right>$.  The $\left|\Psi \right>$ works as an ancilla state, and the sensing information in n dressed states $\left( \left|\phi_n\right> \right)$ is transferred to $\left|\Psi \right>$. For example, $\left|-1, 0 \right>$  of an NV centre can be used for an ancilla state under the irradiation of drive field between $\left|-1, 1 \right>$ and $\left|1, 1 \right>$  of the NV centre described in Fig. 1(a) in the main text.

A minimum detectable value of the magnetic sensor ($B_\mathrm{min}$) is given by the following relation:$B_\mathrm{min} \propto \frac{1}{\sqrt{NT_2}}$,  where $N$ and $T_2$ are number of NV centre (number of qubits) and $T_2$ of NV electron spin~\cite{TaylorNP08S}. In the case of magnetometry with a single NV centre, $N$ = 1.In our experiments, we observed  $T_{2\rho}$ $\sim$ 1.5 ms and $T_2$ $\sim$ 4.2 $\mu$s with and without the generation of dressed states in Fig. 6 of the main paper. It is noted that at least four dressed states were generated in the $T_{2\rho}$ measurement. In this case, we numerically confirm that the sensitivity is approximately enhanced by 27 times using these values and this relation with $N$ = $M$ when 2$M$ virtual states can be prepared by an irradiation of a strong mw-driving field.

\end{document}